\documentclass[prl,twocolumn,amssymb,showpacs]{revtex4}

\usepackage{bm}
\usepackage{graphicx}
\usepackage{dcolumn}
\begin{document}

\title{Dynamics of small perturbations against a stationary polariton
 distribution}
\author{Augusto Gonz\'alez}
\affiliation{Instituto de Cibern\'etica, Matem\'atica y 
 F\'{\i}sica, Calle E 309, Vedado, Ciudad Habana, Cuba}
\pacs{78.20.Bh,71.36.+c,78.47.Cd,42.55.Sf}

\begin{abstract}
The dynamics of small perturbations against the stationary density 
matrix of a pumped polariton system with only one photon polarization
is studied. Depending on the way 
the system is pumped and probed, decay times ranging from 30 to 5000 ps
are found. The large decay times under resonant pumping are related to 
a bottleneck effect in the decay of the excess (probe) populations of 
dark polariton states. No singular behaviour at the threshold for 
polariton lasing is observed.
\end{abstract}

\maketitle

Polariton lasers are lasers without population inversion
\cite{Imamoglu}. Coherence buildup in them is the result of the 
quasibosonic statistics of the polaritons, i.e. quasiparticles composed
by electron-hole pairs from a quantum well strongly interacting with
photons from a semiconductor microcavity. They share similarities with
ordinary photon lasers and Bose-Einstein condensates \cite{Deng,Bajoni}.
In these devices, the threshold power for lasing is 1 - 2 orders of 
magnitude lower than in ordinary lasers. Room-temperature polariton 
lasing has been recently reported \cite{Christopoulos}.

Besides these promising characteristics, polaritons in microcavities
provide an exceptional possibility for fundamental research. An
example is the recent paper [\onlinecite{Ballarini}], where the 
authors seek for evidences of the Goldstone boson, which should appear 
in connection with the buildup of coherence in the polariton system. 
Indeed, in a cylindrical microcavity, the two (approximately) circular
polarizations of the fundamental photon mode are related to two 
degenerate polariton ``condensates''. The relative phase between the
two photon polarizations acts as an order parameter. Phase fixing
leads to a linear polarization \cite{Kavokin}, whose direction can be
easily rotated (the Goldstone excitation). In paper \cite{Ballarini}, 
Ballarini et. al. study the
changes in the PL response of a planar microcavity induced by small 
pulse perturbations of the pumping. They measure the lifetime of these
perturbations, showing that it is much higher than the cavity decay
time, and that it increases when the stationary pumping approaches the
threshold for polariton lasing. These results are interpreted as a
measurement of the lifetime of the Goldstone boson \cite{Carusotto}.

Recent experimental measurements of time-resolved PL in similar systems
\cite{bistability} reveal the importance of both the dynamics involving
a single polarization, and the dynamics involving the two photon
polarizations. On the other hand, a system with a single photon 
polarization - a single polariton condensate - could be realized by
means of a strong magnetic field breaking the degeneracy between the
``right-handed'' and ``left-handed'' condensates.

In the present paper, I consider a model with a single photon polarization
and explore how the way the system is pumped and
probed influences the decay dynamics of the perturbations. I started 
from a scheme, sketched in Refs. 
[\onlinecite{nuestraPRL,distortedGibbs}], in which pumping and photon 
losses in the polariton system are described 
by means of two terms in the master equation for the density matrix.
The linearization of the master equation around the stationary solution 
leads to a system of equations with a source term for the small 
perturbations. The way the system is pumped determines 
the small-oscillation modes of these equations, whereas the way the 
system is probed determines which of the eigenmodes are excited. In an
oversimplified model, with unrealistic parameters, I found decay times
from 30 to 5000 ps, with no singular behaviour at the polariton lasing
threshold. The large decay times correspond to eigenmodes involving
the excitation of dark polariton states which, under resonant pumping,
exhibit a bottleneck effect. 

The starting point is a simple expression for the stationary 
spectral function, $S(\omega)$, describing the PL emission along the cavity
symmetry axis, Eq. (19) of Ref. \onlinecite{laser}:

\begin{equation}
S(\omega)=\frac{1}{\pi}\sum_{I,J}\frac{|\langle I|a|J\rangle|^2
 \rho_{J}^{(\infty)}~\Gamma_{IJ}}{\Gamma_{IJ}^2+(\omega_{IJ}-\omega)^2}.
 \label{eq0}
\end{equation}

\noindent
The magnitudes $\Gamma_{IJ}$ (linewidhts) and $\omega_{IJ}=(E_J-E_I)/\hbar$ 
depend only on the many-polariton wavefunctions and energies, and the 
system parameters, $P$ (pumping rate), and $\kappa=0.1~{\rm ps}^{-1}$ 
(photon losses rate). $\langle I|a|J\rangle$ are the matrix elements of 
the photon annihilation operator between the many-polariton wavefunctions 
$|J\rangle$ and $|I\rangle$. 

In our model, describing a many-exciton quantum dot strongly interacting with
the lowest photon mode of a microcavity \cite{laser}, the wavefunctions and energies,
and from them $\Gamma_{IJ}$, $\omega_{IJ}$,
and $\langle I|a|J\rangle$, are obtained by numerically diagonalizing the
electron-hole-photon Hamiltonian. On the other hand, 
the stationary solutions, $\rho_I^{(\infty)}$ should be obtained from the master
equation for the occupations (coherences are three orders of magnitude 
smaller \cite{distortedGibbs} and will be neglected):

\begin{eqnarray}
\frac{{\rm d}\rho_I}{\rm dt}&=&\kappa\sum_J|\langle I|a|J\rangle|^2 \rho_J
 -\kappa~\rho_I \sum_J |\langle J|a|I\rangle|^2 \nonumber\\
 &+& \sum_{N_{pol}(J)=N_{pol}(I)-1} \rho_J~P_{JI}\nonumber\\
 &-&\rho_I \sum_{N_{pol}(J)=N_{pol}(I)+1} P_{IJ}.
\label{eq2}
\end{eqnarray}

\noindent
$N_{pol}(J)$ is the polariton number (number of electron-hole pairs 
plus number of photons, which is a conserved quantity) of state
$|J\rangle$, and $P_{JI}$ is the pumping rate from state $J$ to state $I$. 
Notice that Eqs. (\ref{eq2}) depends only on $\kappa$, $P$,
and the matrix elements $\langle I|a|J\rangle$. These equations are 
linearly dependent, which expresses the conservation of probability:

\begin{equation}
\sum_I \rho_I = 1.
\label{eq3}
\end{equation}

The stationary solutions, $\rho_I^{(\infty)}$, are obtained by making 
the l.h.s. of Eqs. (\ref{eq2}) equal to zero, and complementing this
homogeneous linear system with the normalization condition, Eq.
(\ref{eq3}). 

Under a small perturbation of the pumping rate, $\delta P(t)$, there is
a variation of the density matrix, $\delta\rho(t)$, and a variation of
the spectral function:

\begin{equation}
\delta S(\omega,t)=\frac{1}{\pi}\sum_{I,J}\frac{|\langle I|a|J\rangle|^2
 ~\delta\rho_{J}(t)~\Gamma_{IJ}}{\Gamma_{IJ}^2+(\omega_{IJ}-\omega)^2}.
 \label{eq1}
\end{equation}

The response, $\delta\rho_I$, to the pulsed perturbation $\delta P(t)$
satisfies the linear system obtained by varying Eqs. (\ref{eq2}):

\begin{eqnarray}
\frac{{\rm d}\delta\rho_I}{\rm dt}&=&\kappa\sum_J \left\{
 |\langle I|a|J\rangle|^2 \delta\rho_J- |\langle J|a|I\rangle|^2 
 \delta\rho_I\right\}\nonumber\\
 &+& \sum_{N_{pol}(J)=N_{pol}(I)-1} \left(\delta\rho_J~P_{JI}
 + \rho_J^{(\infty)} \delta P_{JI}(t) \right) \nonumber\\
 &-& \sum_{N_{pol}(J)=N_{pol}(I)+1} \left( \delta\rho_I P_{IJ}
 +\rho_I^{(\infty)} \delta P_{IJ}(t)\right),
\label{eq4}
\end{eqnarray}

\noindent
which should be complemented with:

\begin{equation}
\sum_I \delta\rho_I = 0.
\label{eq5}
\end{equation}

The structure of Eqs. (\ref{eq4}) is the following: ${\rm d}\delta\rho
/{\rm dt}=A \delta\rho+\delta P(t) B\rho^{(\infty)}$. The eigenvalues
of matrix $A$ are the small oscillation frequencies of the system, 
whereas the probe perturbation, matrix $B$, and the stationary density 
matrix conform the source term determining which oscillation modes are 
excited by the probe pulse.

Let us recall the diagonal terms in Eqs. (\ref{eq4}) in order to get 
a qualitative understanding of the decay times:

\begin{eqnarray}
\frac{{\rm d}\delta\rho_I}{\rm dt}&=&-\kappa \sum_J 
 |\langle J|a|I\rangle|^2 \delta\rho_I \nonumber\\
 &-& \delta\rho_I  \sum_{N_{pol}(J)=N_{pol}(I)+1}P_{IJ} + \cdots
\label{eq6}
\end{eqnarray}

\noindent
This equation shows that the excess occupation of state $I$, $\delta\rho_I$,
can decay in two ways. The first term corresponds to photon emission, and
the second to pumping the excess
occupation to a higher polariton state, $J$, which may further decay through
photon emission. When the state $I$ is dark, $\sum_J |\langle J|a|I\rangle|^2
\approx 0$, only the second mechanism acts. If, in addition, $\delta\rho_I$
is not pumped to higher states because of selective pumping ($P_{IJ}\approx 0$),
then the decay of $\delta\rho_I$ may take very long times. Below, we consider
different regimes of pumping and probing the polariton system.

\begin{figure}[t]
\begin{center}
\includegraphics[width=.95\linewidth,angle=0]{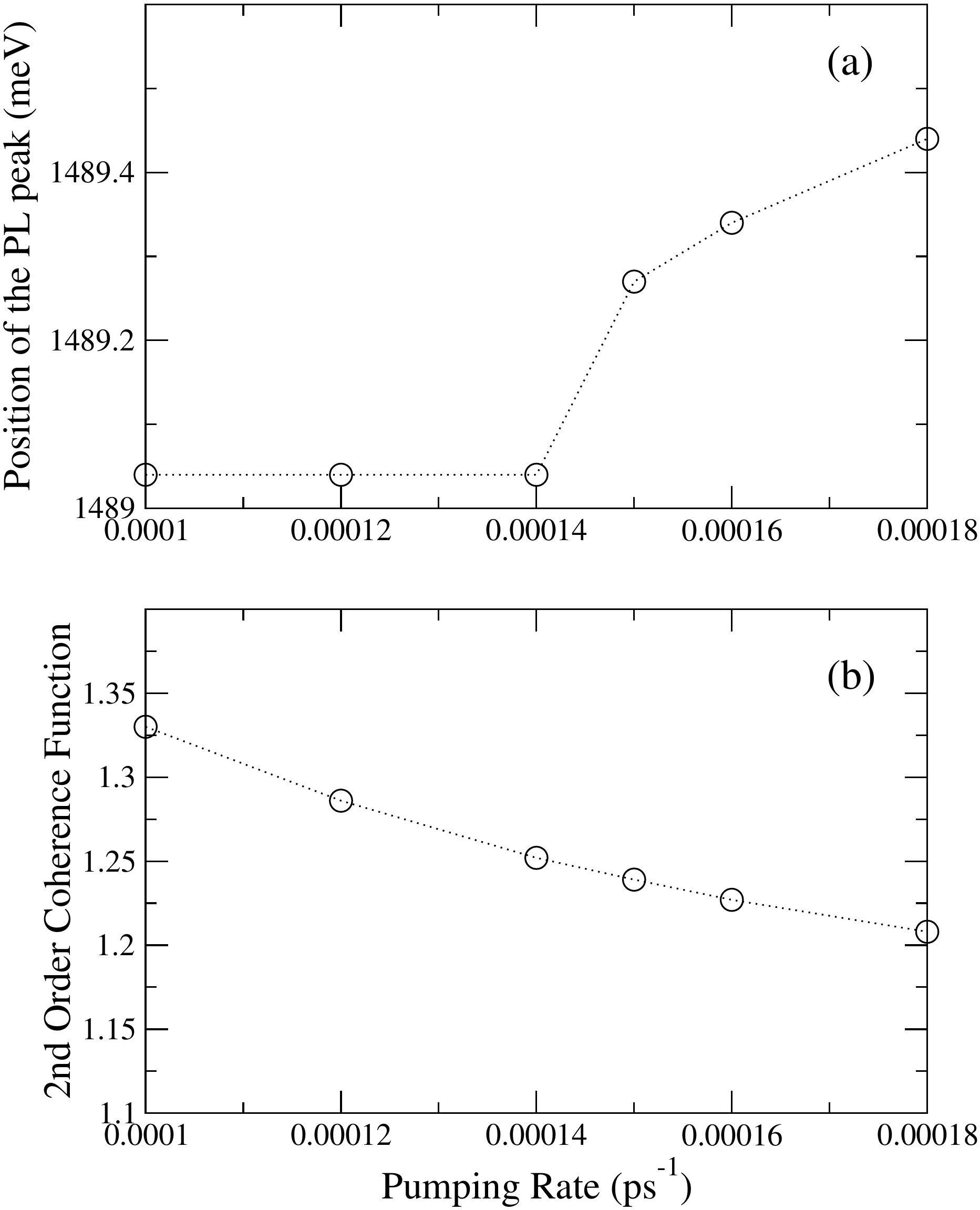}
\caption{\label{fig1} Uniform pumping. (a) Position of the PL line as
a function of the pumping rate. (b) Photon second-order coherence
function.}
\end{center}
\end{figure}

{\bf (a) Uniform pumping and uniform perturbation}

In this case, $P_{IJ}=P$, and $\delta P_{IJ}(t)=\delta P(t)$. This situation
seems to correspond to laser excitation energies well above the lower
polariton branch, and perturbations at these higher energies. 

I show in Fig. \ref{fig1} the position of the main PL line as a function of 
the pumping rate, and the corresponding photon second-order coherence function,
$g^{(2)}(0)$. The jump in the position of the line, and the values near one
of $g^{(2)}(0)$ identify the threshold for polariton lasing at $P_{thr}
\approx 0.00014$ ps$^{-1}$ in the model, where I use the following states
in order to solve Eqs. (\ref{eq2}) for the stationary density matrix: the vacuum 
($I=1$), the 17 existing one-polariton states in the model ($I=2-18$), the 256 
existing two-polariton states ($I=19-274$), and 256 states in each sector with 
$2<N_{pol}\le 10$.

I will study the decay dynamics of the probe for $P$ values in the vicinity of 
$P_{thr}$. The probe pulse is taken in the following way:

\begin{equation}
\delta P(t)=10^{-5}\exp -(t-1)^2~{\rm ps}^{-1},
\label{eq7}
\end{equation}

\noindent
where $t$ is given in ps. We find the $\delta\rho_I$ from Eqs. (\ref{eq4})
and compute, as in the experiment \cite{Ballarini}, the energy-integrated
differential PL response:

\begin{equation}
\delta S(t)=\sum_J \delta\rho_J(t)\sum_I |\langle I|a|J\rangle|^2.
\label{eq8}
\end{equation}

\noindent
The sum over $I$ is restricted to states such that $|E_J-E_I-E_{ref}|<\delta E$,
where $\delta E=2$ meV, and the reference energy in the present case is 
$E_{ref}=1489.2$ meV.

\begin{figure}[t]
\begin{center}
\includegraphics[width=.95\linewidth,angle=0]{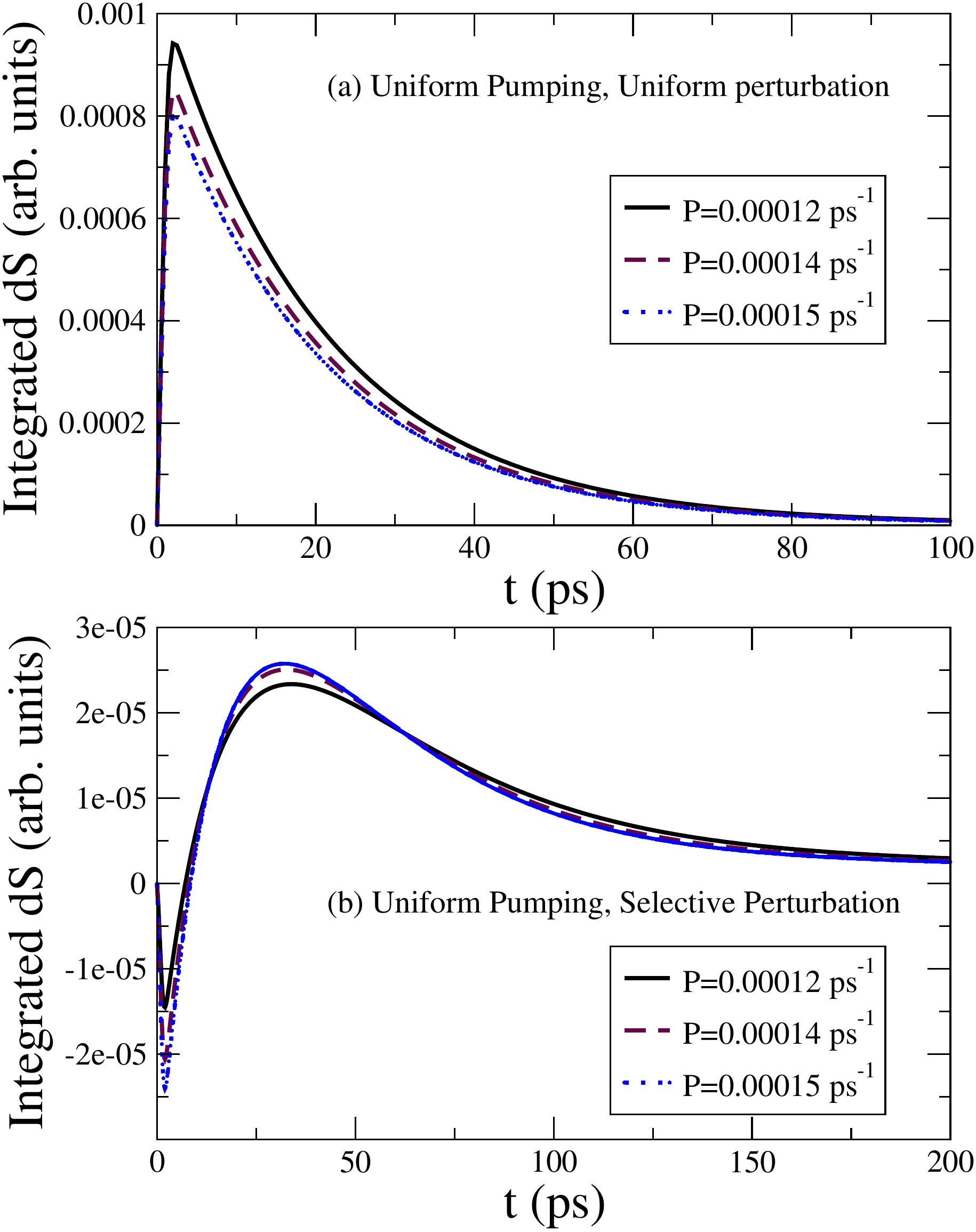}
\caption{\label{fig2} (Color online) Time evolution of the energy-integrated 
differential PL response, Eq. (\ref{eq8}), under uniform pumping. (a) Uniform 
perturbation. (b) Selective perturbation.}
\end{center}
\end{figure}

I draw in Fig. \ref{fig2} (a) the computed $\delta S(t)$ for three values of 
the pumping rate. They show characteristic decay times of around 30 ps. These
values can be understood from the eigenvalues of the linearized decay modes.
Indeed, let us write the smallest (in absolute value) eigenvalues in the 
present case: ..., -0.0358, -0.0287, -0.0117, -0.0067, -0.0023 ps$^{-1}$. 
Notice that they are real, i.e. purely decaying (non propagating) modes. 
The first 
of them (-0.0358 ps$^{-1}$, decay time $\sim 28$ ps, there are many such 
eigenvalues) corresponds basically to an eigenvector in which a single dark 
one-polariton state is excited. The decay is due to pumping to two-polariton 
states with further emission of photons. Notice that $P_{thr}=0.00014$ ps$^{-1}$
times the number of available two-polariton states (256) is equal roughly to the
eigenvalue, -0.0358 ps$^{-1}$. These are the modes dominating the observed 
behaviour of $\delta S(t)$ in the present case. Let us stress that the eigenvalues
practically do not change when $P$ is varied around $P_{thr}$. Thus, we can not
relate the slowest mode (or any other) to the lasing transition. In the scheme of
perturbation I am using, modes with larger decay times are not 
excited. In particular, the last one (-0.0023 ps$^{-1}$, decay time $\sim 434$ ps)
corresponds to an eigenvector involving the simultaneous excitation of one- and
two-polariton dark states combined with higher polariton states. We shall see that
slower decaying modes can be observed by means of a selective excitation.

{\bf (b) Uniform pumping, selective perturbation}

At this point I consider a situation in which the system is pumped at high
energies, as above, but resonantly perturbed. This means that $\delta P_{IJ}(t)$
is given by Eq. (\ref{eq7}), only when the energy difference satisfies the 
inequality $|E_J-E_I-E_{perturb}|<\delta E$. Otherwise it is zero. We show 
results for $E_{perturb}=1499.25$ meV, and  $\delta E=2$ meV. The chosen 
$E_{perturb}$ corresponds to the resonant excitation of a dark one-polariton state,
labeled by the number 15 ($E_{perturb}=E_{15}-E_1$). Of course, other transitions 
may have the same, or close, excitation energy, and could be excited if they 
satisfy the above inequality. Below, we shall discuss in more details, how a dark
state could be perturbed.

The eigenvalues have not changed because I have not modified the pumping scheme.
But the energy-integrated $\delta S(t)$ shows decay times of around 60 ps in the 
present case, indicating the excitation of slower-decaying modes, as compared to 
uniform perturbation. Results are drawn in Fig. \ref{fig2} (b), where an 
oscillation at early times is also observed. It is possible that with a 
different perturbation energy, $E_{perturb}$, or a different perturbation 
strategy the slowest mode could be reached as well.

\begin{figure}[t]
\begin{center}
\includegraphics[width=.95\linewidth,angle=0]{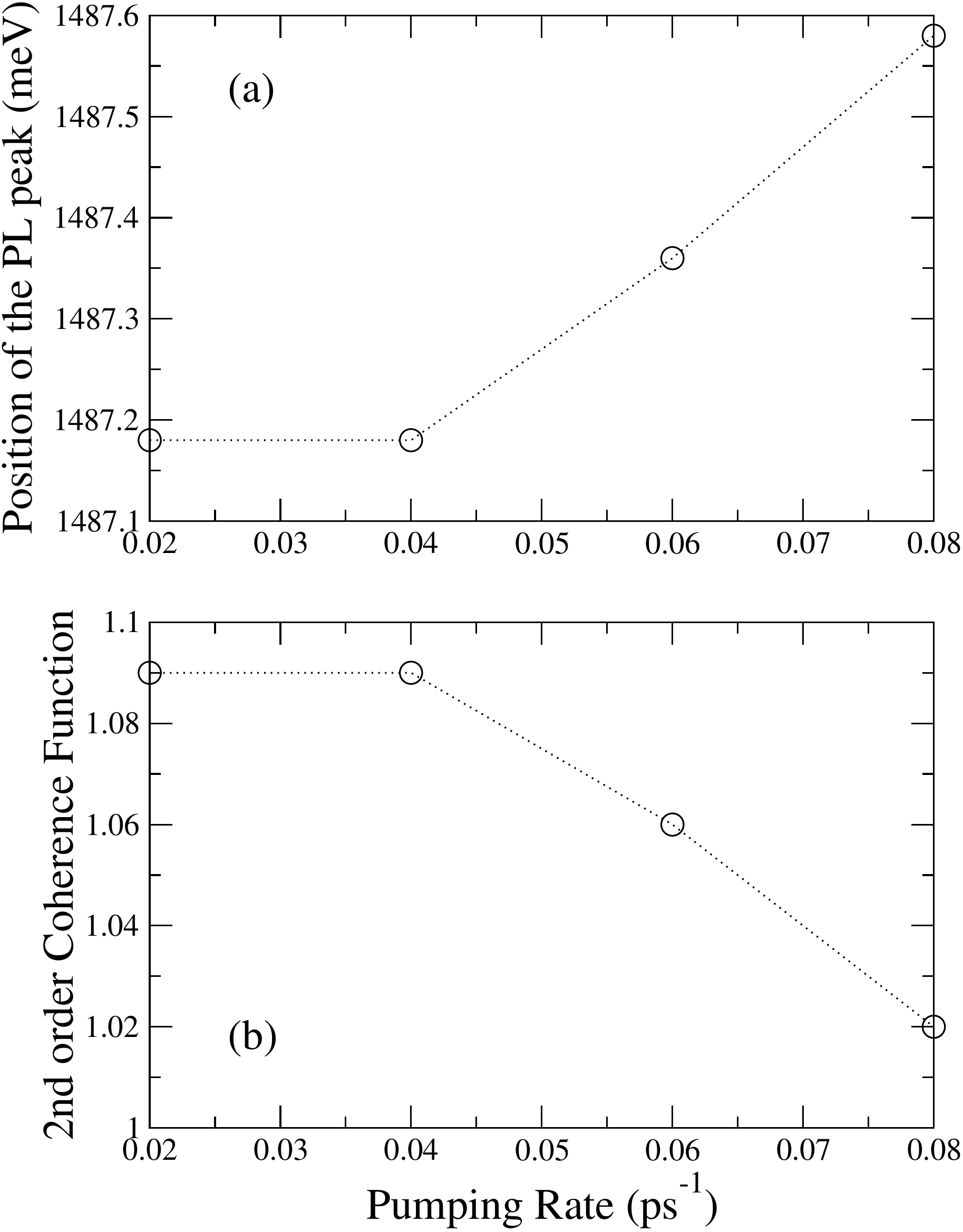}
\caption{\label{fig3} Selective pumping. (a) Position of the PL line as
a function of the pumping rate. (b) Photon second-order coherence
function.}
\end{center}
\end{figure}

{\bf (c) Selective pumping, selective perturbation}

Finally, we consider resonant pumping. Selective pumping can increase
dramatically the decay times because there could be almost dark states, not 
connected to higher states through pumping. The decay of an excess occupation in 
these states could take very long times. 

The stationary pumping rate is chosen in the form: $P_{IJ}=P$ only when 
$|E_J-E_I-E_{pump}|<\delta E$, where $E_{pump}=1488$ meV, and $\delta E=2$ meV. 
The position of the main PL line and the photon second-order coherence function 
are shown in Fig. \ref{fig3}. The jump in the position of the PL line and the 
abrupt variation of $g^{(2)}(0)$ allow us to identify the threshold rate: 
$P_{thr}\approx 0.04$ ps$^{-1}$. Notice that this value is much higher than the 
threshold under uniform pumping, something reasonable. Notice also that even 
below threshold the coherence function take values very close to one. This 
initial coherence is, in some sense, inherited from the pumping.

\begin{figure}[t]
\begin{center}
\includegraphics[width=.95\linewidth,angle=0]{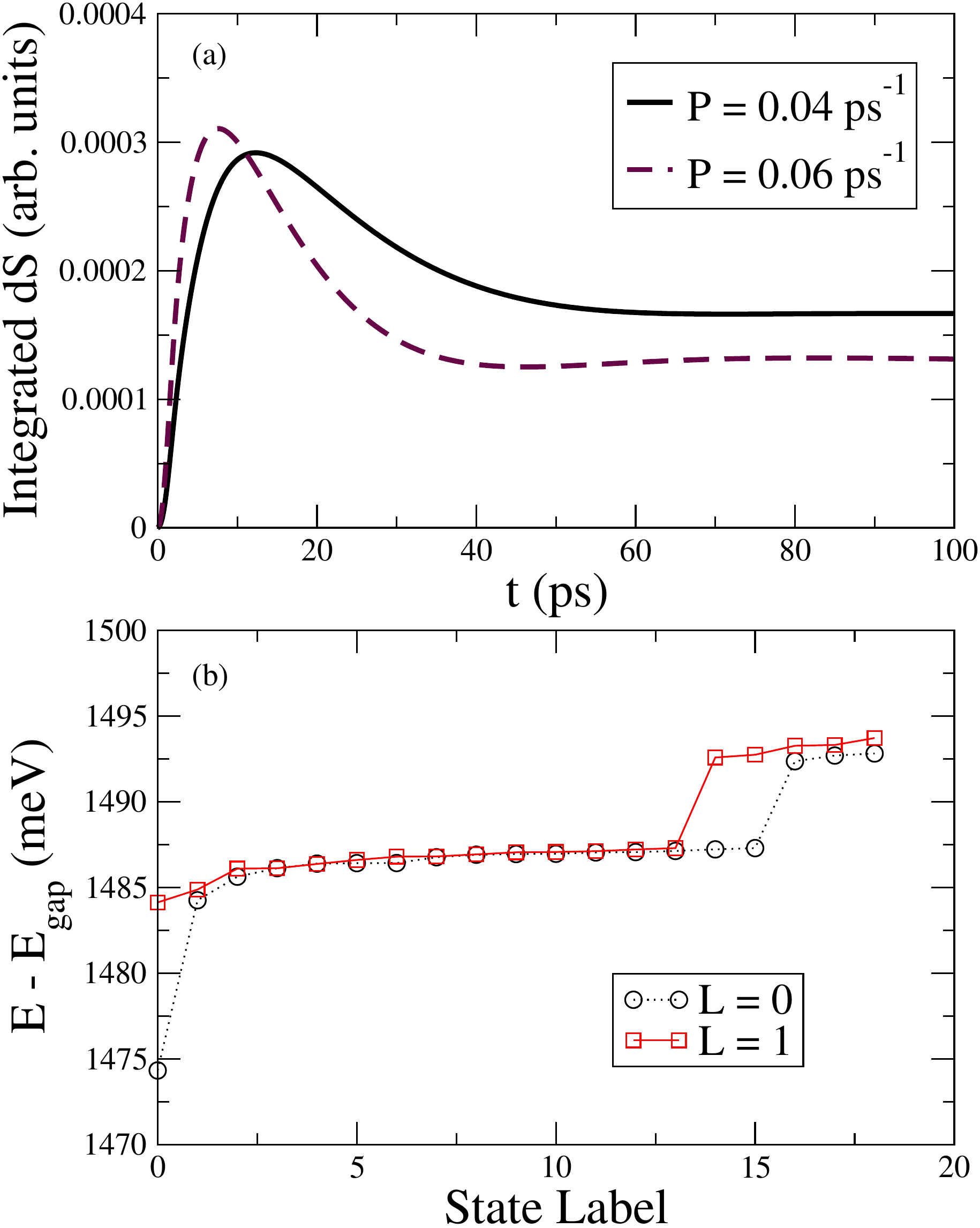}
\caption{\label{fig4} (Color online) (a) Time evolution of the energy-integrated 
differential PL response under selective (resonant) pumping and selective 
perturbation. (b) The lowest 2-polariton states with $L=0$ and $L=1$.}
\end{center}
\end{figure}

The energy-integrated PL response is drawn in Fig. \ref{fig4} (a) for 
$P=0.04$ and $P=0.06$ ps$^{-1}$. After an initial transient period, 
the curves become almost flat, suggesting the excitation of very slow 
decay modes. There are many eigenvalues of the linearized problem with 
absolute value lower than 0.01 ps$^{-1}$ (decay times larger than 100 
ps). The smallest of them is -0.0002 ps$^{-1}$, that is a decay time 
of 5000 ps. Let us stress that no singular behaviour of the eigenvalues 
across the threshold is found, which means that we can not relate any 
of the eigenmodes to the lasing threshold.

Let us consider the question about the selective (resonant) pumping or 
perturbation of a dark state. Electron-hole pairs with a definite energy 
could be injected to the system, but this seems to be difficult to control. 
On the other hand, the direct optical transition is prohibited because, by
definition, the state is dark. However, notice that we are considering 
transitions between $L=0$ states \cite{laser}, which are responsible for
the PL emission along the cavity symmetry axis. States with $L=1$ or
higher, very close in energy to $L=0$ states, are very common, as can be 
seen, for example, 
in Fig. \ref{fig4}(b). These $L=1$ states could be optically excited
(with a non-zero transferred linear momentum, $\vec k$), and then may decay towards
$L=0$ states through emission of very low-energy acoustical phonons.
In the reported experiment \cite{Ballarini}, the system is both pumped and perturbed
by using $\vec k\ne 0$ laser beans.

In conclusion, we studied a model polariton system with a single photon
polarization and computed decay times of probe pulses against a
stationary polariton distribution. Under non-resonant pumping
conditions, the computed decay times are 30-60 ps, whereas under
resonant pumping very large decay times, of the order of thousands of
picoseconds, are obtained. The dark polariton states play a
fundamental role in the decay dynamics, specially under resonant
pumping, where the excess (probe) occupations of particular dark
states may take very long times to decay. No singular behaviour of
any decay mode at the threshold for lasing is observed.

This work was supported by the Programa Nacional de Ciencias 
Basicas (Cuba) and the Caribbean Network for Quantum Mechanics, 
Particles and Fields (ICTP). The author is grateful to Alejandro Cabo
and Alexey Kavokin for discussions.

\end{document}